

\magnification=1200

\parindent=1.5pc
\hsize=6.0truein
\vsize=8.5truein

\centerline{\bf CANCELLATION OF INFRARED DIVERGENCES}
\baselineskip=8pt
\vglue 0.2cm
\centerline{\bf IN QED AT NONZERO TEMPERATURE}
\vglue 0.4cm
\centerline{ H. ARTHUR WELDON}
\baselineskip=13pt
\centerline{\it  Department of Physics, West Virginia University}
\baselineskip=12pt
\centerline{\it Morgantown, WV 26506-6315, USA}
\vglue 0.4cm
\centerline{Talk presented at the 3rd Workshop on Thermal Field Theories}

\centerline{ and their Applications, Banff, Alberta (August, 1993)}

\vglue 0.6cm
\centerline{\tenrm ABSTRACT}
\vglue 0.3cm
{\rightskip=3pc
 \leftskip=3pc
 \tenrm\baselineskip=12pt\noindent
 The radiation produced by a classical
charged current coupled to a quantized $A_{\mu}$ is first computed. To each
order in $\alpha$, all
infrared divergences cancel between the virtual $\gamma$'s and the real
$\gamma$'s that are absorbed
from or emitted into the plasma.  When all
orders of perturbation theory are summed, the finite answer predicts a
suppression of radiation with
$\omega< \alpha T$. The analysis of QED then consists of two steps. First, a
general probability at
$T\neq 0$ is organized so that all the virtual $e^{\pm},\gamma$ are in the
amplitudes  and all the
real $e^{\pm},\gamma$ are in the phase space
integrations. Next, the cancellations of IR divergences
between virtual and real are demonstrated.
\vglue 0.4cm}
\vfil
\baselineskip=14pt
\leftline{\bf 1. Introduction}
\vglue 0.3cm
Field theories containing massless particles contain infrared divergences,
which
 may be classified into three types as follows.

{\it Soft or infrared divergences:} When
an on-shell electron ($p^{2}=m^{2}$) emits an on-shell photon ($k^{2}=0$) the
resulting electron
is off-shell with a propagator   $${-1\over (p-k)^{2}-m^{2}}={1\over 2p\cdot
k}=
{1\over 2(E-|\vec{p}|\cos\theta)|\vec{k}|}\eqno{(1)}$$ Since this
emission amplitude  is proportional to $1/k$, the emission rate is
$$\int {d^{3}k\over 2k}\ |{1\over k}|^{2}\ [1+n_{B}(k)]\hskip1cm n_{B}(k)\equiv
{1\over \exp(k/T)-1}
\eqno{(2)}$$
At $T=0$ this gives a  logarithmic divergence at small $k$ but the divergence
cancels in physical
quantities.$^{1}$ At $T\neq 0$  the divergence is linear at small $k$
because of the
Bose-Einstein function  $n_{B}$. This paper will sketch how the linear and
logarithmic divergences
all cancel in physical quantities.

{\it Mass or collinear divergences:} If the electron mass were zero in Eq. (1),
then
$E=p$ and there would be a singularity in the amplitude at $\theta=0$ for any
value of $k$.
These collinear singularities beset nonabelian theories but are not present in
massive QED, which
is the subject here.

{\it Coulomb divergences:} The total cross section for charged particle
scattering
is divergent
$$\sigma_{\rm Tot}\sim\int {\sin\theta\ d\theta\over \sin^{4}\theta}
\sim \int {dt\over t^{2}}\sim \infty\eqno{(3)}$$
At $T=0$ this is not too troublesome. At $T\neq 0$ it arises in
self-energy discontinuities at two loops as can be seen even in the classical
expression
$\Gamma=nv\sigma_{\rm Tot}$ for the collision rate. At $T\neq 0$  this
divergence
is reduced to logarithmic by Braaten-Pisarski resummation$^{2}$ but it is not
eliminated.

\vglue 0.6cm
\leftline{\bf 2. IR Cancellation  for  Semiclassical Bremsstrahlung}
\vglue 0.4cm
\leftline{\it 2.1. Bremsstrahlung to First Order in $\alpha$}
\vglue1pt
The first step is to examine the radiation produced by
 a charged particle that scatters while passing through a fixed-temperature
plasma.
To first order in $\alpha$ the probability of radiating energy
$\omega=|\vec{k}|$ is
$$2\omega{dP_{1}\over d^{3}k}=\sum_{\rm pol}|{\cal M}(k)|^{2}
\ [1+n_{B}(\omega)]/(2\pi)^{3}\eqno{(4)}$$
When  $\omega$ is small,  the matrix element is
$${\cal M}(k)=J^{\mu}(k)\epsilon_{\mu}=ie\left({p^{\prime}\cdot\epsilon\over
p^{\prime}\cdot k}-
{p\cdot\epsilon\over p\cdot k}\right)
\ e^{-k/2\Lambda}\eqno{(5)}$$
regardless of the spin of the charged particle.
 Here $\Lambda$ is a momentum cutoff that
is necessary later.  The radiation is mostly parallel to $\vec{p}$ or
$\vec{p}^{\ \prime}$.
When integrated  over angles the result is
$${dP_{1}\over d\omega}={A\over\omega} [1+n_{B}(\omega)]\ e^{-\omega/2\Lambda}
\hskip0.5truein
A(p\cdot p^{\prime})
\equiv {\alpha\over \pi}\left[{1\over v}\ln({1+v\over
1-v})-2\right]\eqno{(6)}$$
 where $v$  is defined by $p\cdot p^{\prime}=m^{2}(1-v^{2})^{-1/2}$.  Except
for the statistical
factor $n_{B}$, Eq. (6)
 is a classical formula. When $\omega\ll T$ it predicts
$dP_{1}/ d\omega\approx AT/\omega^{2}.$
This is totally unphysical because the   energy radiated in low energy modes,
below some $E_{\rm
max}$, would be infinite:  $\int_{0}^{E_{\rm max}} d\omega
\omega dP_{1}/d\omega=\infty.$
\vglue0.3cm
\noindent {\it 2.2 Bremsstrahlung to All Orders in $\alpha$}
\vglue1pt
To improve on the first-order result, we  couple the classical current in Eq.
(5) to the quantized
radiation field $A_{\mu}$. The generating  function for all  multi-photon
amplitudes can be obtained
by a  functional integration over $A_{\mu}$. From this one obtains the
multi-photon amplitudes
${\cal M}$.  The probability that any number $n$ of real photons in the plasma
will radiate a
\underbar{net} energy $\omega$  is $${dP\over d\omega}=\sum_{n=1}^{\infty}\int
d\Phi_{1}...d\Phi_{n}
\delta(k_{1}^{0}+...k_{n}^{0}-\omega)
{1\over n!}\sum_{\rm pol}|{\cal M}(k_{1},...k_{n})|^{2}\eqno{(7)}$$
$$d\Phi_{i}\equiv{d^{4}k_{i}\over
(2\pi)^{3}}\delta(k_{i}^{2})\
[\theta(k_{i}^{0})+n_{B}(|\vec{k}_{i}|)]\eqno{(8)}$$
This weights photon emission ($k^{0}>0$)  with the statistical factor
$1+n_{B}$; and
photon absorption ($k^{0}<0$)  with the factor $n_{B}$.

 Each amplitude ${\cal M}$ is infrared divergent from closed loops of virtual
photons.
  Each  integration $d\Phi$ over the real photons is also infrared divergent.
The virtual and real contributions exponentiate to give
$${dP\over d\omega}=\int_{-\infty}^{\infty}{dz\over 2\pi}e^{-i\omega z}
\exp[\overline{R}(z)]\eqno{(9)}$$
$$\overline{R}(z)
\equiv\int{d^{3}k\over 2k(2\pi)^{3}}J_{\mu}(k)J^{\mu}(k)
\left([1+n_{B}]e^{ikz}
+n_{B}e^{-ikz}-[1+2n_{B}]\right)\eqno{(10)}$$
Real emission, real absorption, and virtual photons are automatically weighted
by
$1+n_{B}$, $n_{B}$, and $1+2n_{B}$, respectively.
Even though  $J_{\mu}\sim 1/k$ and $n_{B}\sim 1/k$ at small $k$,
 Eq. (10) is competely finite in the region $k\to 0$.
 It is possible to  compute
$\overline{R}(z)$ and to compute the Fourier transform $dP/d\omega$. The final
result$^{3}$ is
$${dP\over d\omega}=|\Gamma({A\over 2}+i{\omega\over 2\pi T})|^{2}\
{e^{\omega/2T}e^{-|\omega|/\Lambda}\over 4\pi^{2}T\ \Gamma(A)}\ \left({2\pi
T\over
\Lambda}\right)^{A}\eqno{(11)}$$
where $A$ is given in Eq. (6). The most interesting feature  of Eq. (11) is the
appearance of two dimensionless scales:  $A\ll 1$ and $\omega/T$. If $A\pi
\ll\omega/T$ then
$${dP\over d\omega}\approx{A\over\omega} [1+n_{B}(\omega)]\ e^{-\omega/\Lambda}
\hskip0.3truein (A\pi T\ll\omega)\eqno{(12)}$$
which coincides with the first-order result Eq. (6).
However this does not apply  at $\omega\ll T$. At small energy,
$${dP\over d\omega}\approx{AT\over \omega^{2}+(A\pi T)^{2}}\hskip1cm
(\omega\ll T)\eqno{(13)}$$
Naturally Eq. (12) and (13) agree  in the region of overlap. At very small
energies,
$\omega\ll A\pi T$, $dP/d\omega$ is
constant  rather than increasing like $1/\omega^{2}$.
The  interpretation of the small $\omega$ suppression is that the quantity $2
A\pi T=\Gamma_{r}$ is a
damping rate produced by the radiation reaction that is required by
unitarity$^{3}$.

\vglue 0.6cm
\leftline{\bf 3. IR Cancellation for Thermal QED}
\vglue 0.4cm
\leftline{\it 3.1 Separation of Real and Virtual Particles}
\vglue0.3cm
We now set aside the semiclassical approximation and turn to the full quantum
field theory.
Each species of particle has four types of propagator in thermal field theory:
$S_{ab}$ for electrons and $D_{ab}^{\mu\nu}$ for photons  with $a,b=1\ {\rm
or}\ 2$.
The problem is that $D_{12}=D_{21}\sim\delta(k^{2})$ and
$S_{12}=-S_{21}\sim\delta(p^{2}-m^{2})$
actually represent real, on-shell particles but this is obscured by the
finite-temperature
 Feynman rules. Before looking for infrared cancellations, it is very
convenient to
reorganize thermal probabilities as squares of amplitudes that contain only
$S_{11}$ and
$D_{11}^{\mu\nu}$, integrated over physical phase space.

For definiteness, consider the process $e^{-}(p_{1})$ + plasma $\to$  anything,
with rate
$$R(p_{1})=\sum_{F,I}|<F|C|I>|^{2}\hskip 1cm
C\equiv[S,b^{\dagger}(\vec{p}_{1})]\eqno{(14)}$$
Using completeness and thermofield dynamics$^{4}$, one can write this as
$$R(p_{1})=<0(\beta)|C^{\dagger}C|0(\beta)>=\sum_{F}|<F(\beta)|C|0(\beta)>|^{2}\eqno{(15)}$$
where $|F(\beta)>$ is a complete set of thermal states in the Foch space built
out of
Bogoliubov-transformed creation operators$^{4}$. For example, the contribution
of
 $a_{\beta}^{\dagger}|0(\beta)>$  to $R$ is
$$\int {d^{3}k\over 2k(2\pi)^{3}}[1+n_{B}(k)]\
|<0(\beta)|[a,C]|0(\beta)>|^{2}\eqno{(16)}$$
The contribution of  $\tilde{a}_{\beta}^{\dagger}|0(\beta)>$  to $R$ is
$$\int {d^{3}k\over 2k(2\pi)^{3}}n_{B}(k)\
|<0(\beta)|[a^{\dagger},C]|0(\beta)>|^{2}\eqno{(17)}$$
The complete rate is
$$R(p_{1})=\sum_{\ell=2}^{\infty} \int d\Psi_{2}...d\Psi_{\ell}
\sum_{n=0}^{\infty}\int d\Phi_{1}...d\Phi_{n}\
{1\over n!}|M_{\ell,n}(p_{1},...p_{\ell};k_{1},...k_{n})|^{2}\eqno{(18)}$$
where $d\Phi$ is the photon phase space  from Eq. (8),
$d\Psi$ is the fermion phase space, and
${\cal M}_{\ell n}$ is the amplitude for $\ell$ real
leptons (including the observed $p_{1}$) and
 $n$ real photons (either initial or final).  The amplitude is
$$M_{\ell,n}=<0(\beta)|[a^{\#}(\vec{k}_{n}),[a^{\#}(\vec{k}_{n-1}),...\{b^{\#}(\vec{p}_{2}),
[b^{\dagger}(\vec{p}_{1}),S]\}...]]|0(\beta)>\eqno{(19)}$$
In the interaction picture, the averaging required for these amplitudes is
determined by
the operator $S\exp(-\beta H)$ rather than by $\exp(-\beta H)$. Consequently
these amplitudes contain only the propagators $D^{\mu\nu}_{11}$ and $S_{11}$
and none of the cut
propagators $D^{\mu\nu}_{12}$ and $S_{12}$.
 Eq. (18) therefore provides a complete separation between real particles
(contained in the phase
space $d\Psi d\Phi$) and virtual particles (represented by the propagators
$D_{11}$ and $S_{11}$).
Ref. 5 and 6 present diagramatic arguments for this result.

\vglue0.3cm
\leftline{\it 3.2 IR Cancellation to All Orders in $\alpha$}
\vglue0.3cm
To analyze the rate in Eq. (18) one can repeat the  analysis of Yennie,
Frautschi, and  Suura$^{1}$
with  some modifications.
For fixed fermion
momenta, define
  $$R_{\ell}(p_{1},...p_{\ell})=\sum_{n=0}^{\infty}\int
d\Phi_{1}...d\Phi_{n}{1\over
n!}|M_{\ell,n}|^{2}\eqno{(20)}$$
Virtual photons inside the amplitude ${\cal M}_{\ell n}$  can only cause an IR
divergence when
they are on-shell and attached to ``external" fermions $p_{1},...p_{\ell}$.
Each end of a photon line attached to an ``external" fermion gives a
multiplicative factor
$ e2p^{\mu}/ 2p\cdot k$ plus non-IR terms.
 At each order of $\alpha$, ${\cal M}_{\ell n}$ has a maximum IR
divergence $\alpha^{p}$ plus non-leading divergences $\alpha^{p-1}$,
$\alpha^{p-2}$,....
 When summed to all orders, the leading and
 non-leading divergences from virtual photons all exponentiate.

Next we examine infrared divergences produced by real photon emission and
absorption.
These arise from  the integrations $d\Phi$ in Eq. (20) whenever ${\cal M}_{\ell
n}\sim 1/k$.  The
diagrams in which $n$ real photon lines are attached to the ``external"
fermions in all possible
ways,  give multiplicative factors $ e2p^{\mu}/2p\cdot k$ plus non-IR terms.
The contribution to Eq. (20) of $n$ real photons
has an IR divergence $\alpha^{n}$ plus non-leading divergences $\alpha^{n-1}$,
$\alpha^{n-2}$, .. 1.
When summed  over $n$, the leading and non-leading
 divergences from real photons also exponentiate. The final result is
 $$R_{\ell}(p_{1},...p_{\ell})=\sum_{m=0}^{\infty}\int_{-\infty}^{\infty}
d\omega{dP\over d\omega}
\int d\Phi_{1}...d\Phi_{m}
\beta_{\ell,m}(k_{1},...k_{m})/m!\eqno{(21)}$$
with all IR divergences  contained in the quantity
$${dP\over d\omega}=\int_{-\infty}^{\infty}{dz\over 2\pi}e^{-i\omega z}
\exp[\overline{R}(z)]\eqno{(22)}$$
$$\overline{R}(z)
=\int{d^{3}k\over 2k(2\pi)^{3}}\sum_{a,b=1}^{\ell}{e_{a}e_{b}(p_{a}\cdot
p_{b})\over (p_{a}\cdot k)
(p_{b}\cdot k)}
\left([1+n_{B}]e^{ikz}
+n_{B}e^{-ikz}-[1+2n_{B}]\right)e^{-k/\Lambda}\eqno{(23)}$$
This integration is IR finite because of the delicate cancellation between
real emission ($1+n_{B}$), real absorption ($n_{B}$), and virtual photons
($1+2n_{B}$).
 The integrations are
the same as in Sec. 2 with the result $${dP\over d\omega}=|\Gamma({A\over
2}+i{\omega\over 2\pi
T})|^{2}\ {e^{\omega/2T}e^{-|\omega|/\Lambda}\over 4\pi^{2}T\Gamma(f)}({2\pi
T\over
\Lambda})^{A}\eqno{(24)} $$
but now $A$ depends on all the ``external" fermion momenta:
$$A(p_{1},...p_{\ell})\equiv
-\sum_{a,b=1}^{\ell}{e_{a}e_{b}\over 8\pi^{2}}{1\over v_{ab}}
\ln\left({1+v_{ab}\over
1-v_{ab}}\right)\quad \ge 0\eqno{(25)}$$ and $v_{ab}$, defined by $p_{a}\cdot
p_{b}/m^{2}=(1-v_{ab}^{2})^{-1/2}$,
 is relative velocity of charge $a$ in the rest frame of  $b$.

Eq. (24) is infrared finite and has the behavior
$dP/d\omega\to$ constant as $\omega\to 0$.
Therefore, for fixed momenta of the ``external" fermions, the rate  in Eq. (21)
contains no infrared divergences.

\vglue 0.6cm
\leftline{\bf 4. Conclusions}
\vglue 0.4cm
Although the infrared divegences have all been eliminated, the analysis is not
quite
complete.
 The full rate requires integration  over the
 thermalized fermions:
$$R(p_{1})=\sum_{\ell=2}^{\infty} \int
d\Psi_{2}...d\Psi_{\ell}\ R_{\ell}(p_{1},...p_{\ell})\eqno{(26)}$$
The fermion integrations do not affect the infrared finiteness of $R(p_{1})$.
However,
there will still be Coulomb divergences  that arise when any one of the
momentum transfers
vanishes: $(p_{a}-p_{b})^{2}\to 0$. It is known that the usual zero-temperature
Coulomb divergence
$\int \sin\theta d\theta/\theta^{4}$ is reduced to logarithmic, $\int
\sin\theta d\theta/\theta^{2}$
at $T\neq 0$ due to Braaten-Pisarski resummation$^{2}$. These logarithmic
divergences are not
elimitated.

For definiteness, the process
$e^{-}(\vec{p}_{1})$ + plasma $\to$ anything was treated specifically.
One can analyze a general process $\{ A\}$ + plasma $\to \{B\}$ + anything,
where $\{A\}$ and $\{B\}$ are any sets of $e^{\pm}, \gamma$ by using the same
amplitudes
${\cal M}_{\ell,n}$. The rate corresponding to Eq. (18) would not be integrated
over the
lepton and photon momenta that enter and leave the plasma. These fixed external
momenta
cause no complication, but make the notation a bit more cumbersome.
The same arguments apply and show that
all infrared divergences cancel in the generalized rates. However, the
logarithmic Coulomb divergence
remain.
\vglue 0.6cm
\leftline{\bf 5. Acknowledgements}
\vglue 0.4cm
It is a pleasure to thank R. Kobes and G. Kunstatter for organizing an
excellent workshop.
This work was supported in part by the U.S. National Science Foundation under
grant PHY-9213734.

\vglue 0.6cm
\leftline{\bf 6. References}
\vglue 0.4cm

\medskip
\itemitem{1.} D.R. Yennie, S.C. Frautschi, and H. Suura, {\it Ann. Phys. (NY)}
{\bf 13} (1961) 379.
\itemitem{2.} R.D. Pisarski, {\it Phys. Rev. Lett. } {\bf 63}  (1989) 1129;
E. Braaten and R.D. Pisarski, {\it Phys. Rev.}{\bf 64}  (1990) 1338.
\itemitem{3.} H.A. Weldon, to appear in {\it Phys. Rev.} {\bf D}.
\itemitem{4.} H. Matsumoto, I. Ojima, and H. Umezawa, {\it Ann. Phys. (NY)}
 {\bf 152} (1984) 348.
\itemitem{5.} N. Ashida, H. Nakkaggawa, A. Ni\'egawa, and H. Yokata, {\it Phys.
Rev.}
{\bf D45} (1992) 2006; {\it Ann. Phys. (NY)} {\bf 215} (1992) 315.
\itemitem{6.} A. Ni\'egawa and K. Takashiba, {\it Nucl. Phys.} {\bf B370}
(1992) 335.

\end